\documentclass[10pt,a4paper]{article}
\usepackage{amsmath}
\usepackage{amsfonts}
\usepackage{amssymb}
\usepackage{color}
\usepackage[left=2cm,right=1cm,top=1.5cm,bottom=1.5cm]{geometry}
\usepackage{hyperref}

\newtheorem{Proposition}{Proposition}
\newtheorem{Corollary}{Corollary}
\newtheorem{Theorem}{Theorem}
\input{tcilatex}
\begin{document}

\title{Evaluating the Maximal Violation of the Original Bell Inequality by
Two-Qudit States Exhibiting Perfect Correlations/Anticorrelations\bigskip }
\author{Andrei Y. Khrennikov $^{1}$ and Elena R. Loubenets $^{2}\bigskip $ \\
$^{1}$International Center for Mathematical Modeling, Linnaeus University,
35195, Vaxjo, Sweden\medskip \\
$^{2}$National Research University Higher School of Economics, 101000
Moscow, Russia}
\maketitle

\begin{abstract}
We introduce the general class of symmetric two-qubit states guaranteeing
the perfect correlation or anticorrelation of Alice and Bob outcomes
whenever some spin observable is measured at both sites. We~prove that, for
all states from this class, the~maximal violation of the original Bell
inequality is upper bounded by $\frac{3}{2}$ and specify the two-qubit
states where this quantum upper bound is\ attained. The~case of two-qutrit
states is more complicated. Here,~for~all two-qutrit states, we~obtain the
same upper bound $\frac{3}{2}$\ for violation of the original Bell
inequality under Alice and Bob spin measurements, but~we have not yet been
able to show that this quantum upper bound is the least one. We~discuss
experimental consequences of our mathematical study.

\end{abstract}

\section{Introduction}

The recent loophole free experiments~\cite{B1, B2, B3} demonstrated
violations of classical bounds for the wide class of the Bell-type
inequalities which derivations are not based on perfect (anti-)
correlations, for~example, the~Clauser--Horne--Shimony--Holt (CHSH)
inequality ~\cite{CHSH} and its further various generalizations~\cite{CHSH1,
CHSH2, CHSH3, CHSH4, B11, B22, B33, B4, B5, B6}. These~experiments have very
high value for foundations of quantum mechanics (QM) and interrelation
between QM and hidden variable models, see, for~example,~\cite{A1, A2, A3,
A4, OPH, Cour, R3, R4}~for recent debates.

However, {John Bell}~started his voyage beyond QM not with such
inequalities, but with the {original Bell inequality}~\cite{L1, L2} the
derivation of which is based on perfect anticorrelations---the condition
which is explicitly related to the Einstein--Podolsky--Rosen (EPR) argument~%
\cite{EPR}.

At the time of the derivation of the original Bell inequality, the
experimental technology was not so advanced and preparation of sufficiently
clean ensembles of singlet states was practically dificult. Therefore, Bell
enthusiastically supported the proposal of Clauser, Horne, Shimony, and
Holt, which is based on a new scheme (without exploring perfect
correlations) and the CHSH inequality~\cite{CHSH}.

The tremendous technological success of recent years, especially, in
preparation of the two-qubit singlet state and high efficiency detection,
makes the original Bell's project at least less difficult. This~novel
situation attracted again attention to the original Bell inequality~\cite{L8}%
. We~also point to related theoretical studies on the original Bell
inequality which were done during the previous years, see~\cite{ANT, L, L52,
L5, L3}. In~\cite{L52, L3}, it~is, for~example, shown that, unlike the CHSH\
inequality, the~original Bell inequality distinguishes between classicality
and quantum separability.

Finally, we~point to a practically unknown paper of Pitowsky~\cite{L7} where
he claims that by violating the original Bell inequality and its
generalizations it would be possible to approach a higher degree of
nonclassicality than for the CHSH-like inequalities.

This claim is built upon the fact that, for~the CHSH inequality $\left\vert 
\mathcal{B}_{clas}^{\mathrm{CHSH}}\right\vert \leq 2,$ the fraction $F_{%
\mathrm{CHSH}}^{(\rho _{d})}$\ of the quantum (Tsirelson) upper bound~\cite%
{L9, L10} $2\sqrt{2}$\ to the classical one is equal to $F_{\mathrm{CHSH}%
}^{(\rho _{d})}=\sqrt{2}$\ for a bipartite state $\rho _{d}$\ of an
arbitrary dimension $d\geq 2$, whereas, for~the original Bell inequality,
the fraction $F_{\mathrm{OB}}^{(\rho _{singlet})}$\ of the quantum upper
bound for the two-qubit singlet\ ($d=2$) to the classical bound ({equal to
one} see in Section~\ref{s2}) is given by~\cite{L7, L8}

\begin{equation}
F_{\mathrm{OB}}^{(\rho _{singlet})}=\frac{3}{2}>\sqrt{2}=F_{\mathrm{CHSH}%
}^{(\rho _{d})},\text{ \ \ }\forall d\geq 2.  \label{OB1}
\end{equation}

The rigorous mathematical proof of the least upper bound $\frac{3}{2}$\ on
the violation of the original Bell inequality by the two-qubit singlet was
presented in the article~\cite{L8} written under the influence of Pitowsky's
paper~\cite{L7}. In~both papers---References~\cite{L8, L7},
the~considerations were restricted only to the two-qubit singlet case.

However, for~the violation $F_{\mathrm{OB}}^{(\rho _{d})}$\ of the original
Bell inequality by a two-qudit state $\rho _{d}$ exhibiting perfect
correlations/anticorrelations, the~CHSH inequality implies for all $d\geq 2$
the upper bound $(2\sqrt{2}-1)$ (see in Section~\ref{s3}) and the latter
upper bound is more than the least upper bound $\frac{3}{2}$\ proved~%
\mbox{\cite{L7, L8}} for the two-qubit singlet.

We stress that quantum nonlocality is not equivalent~\cite{L6} to quantum
entanglement and that larger violations of Bell inequalities can be reached ~%
\cite{Carlos} by states with less entanglement. Therefore,~the~proof~\cite%
{L8} that, for~the two-qubit singlet state (which is maximally entangled),
the least upper bound on violation of the original Bell inequality is equal
to $\frac{3}{2}$ does not automatically mean that $\frac{3}{2}$\ is the
least upper bound on violation of the original Bell inequality for all
two-qubit states. Moreover,~the~proof of the least upper bound $\frac{3}{2}$%
\ on violation of the original Bell inequality by the singlet state has no
any consequence for quantifying violation of this inequality by a two-qudit
state of an arbitrary dimension $d\geq 2.$

In the present paper, we~rigorously prove that under Alice and Bob spin
measurements, the~least upper bound $\frac{3}{2}$\ on the violation of the
original Bell inequality holds for all two-qubit and all two-qutrit states
exhibiting perfect correlations/anticorrelations. In~the sequel to this
article, we~intend to prove that, quite similarly to the CHSH case where the
least upper bound $\sqrt{2}$ on quantum violations holds for all dimensions $%
d\geq 2,$ {under the condition on perfect} correlations/anticorrelations,
the~least upper bound $\frac{3}{2}$\ on quantum violations of the original
Bell inequality holds for all $d\geq 2$ (see in {Section} \ref{s6}).

In Section~\ref{s2} (Preliminaries), we~present the condition~\cite{L3} on
perfect correlations or anticorrelations for joint probabilities and prove,
under this condition, the validity of the original Bell inequality in the
local hidden variable (LHV)\ frame. This~general condition is true for any
number of outcomes at each site and reduces to the Bell's perfect
correlation/anticorrelation condition~\cite{L1} on the correlation function
only in case of Alice and Bob outcomes $\pm 1.$

In Section~\ref{s3}, we~analyse violation of the original Bell inequality by
a two-qudit quantum state and show that, for~all dimensions of a two-qudit
state exhibiting perfect correlations/anticorrelations and any three qudit
observables, the~maximal violation of the original Bell inequality cannot
exceed the value $(2\sqrt{2}-1).$\ 

In Section~\ref{s4}, we~introduce (Proposition \ref{p2}) the general class
of symmetric two-qubit density operators which guarantee perfect correlation
or anticorrelation of Alice and Bob outcomes whenever some (the same) spin
observable is measured at both sites. We~prove (Theorem \ref{theorem 1})
that, for~all states from this class, the~maximal violation of the original
Bell inequality is upper bounded by $\frac{3}{2}$\ and specify the two-qubit
states for which this quantum upper bound is attained.

In Section~\ref{s5}, we~consider Alice and Bob spin measurements on
two-qutrit states. This~case is more complicated. Here,~we~are also able to
prove the upper bound $\frac{3}{2}$\ for all spin measurements on an
arbitrary two-qutrit state, but~we have not yet been able to find two-qutrit
states for\ which this upper bound is attained. In~future, we~plan to study
this problem as well as to consider spaces of higher~dimensions.

In {Secton} \ref{s6}, we~summarize the main results and stress that
description of\ general density operators ensuring perfect correlations or
anti-correlations for spin or polarization observables may simplify
performance of a hypothetical experiment on violation of the original Bell
inequality. In~principle, experimenters need not prepare an ensemble of
systems in the singlet state since, by~Proposition \ref{p2} and Theorem \ref%
{theorem 1}, for~such experiments, a~variety of two-qubit states, pure and
mixed, can~be used and it might be easier to prepare some of such states.

\section{Preliminaries: Derivation of the Original Bell Inequality in a
General Case \label{s2}}

Both Bell's proofs~\cite{L1, L2} of the original Bell inequality in a local
hidden variable (LHV) frame are essentially built up on two assumptions: a
dichotomic character of Alice's and Bob's measurements plus the perfect
correlation or anticorrelation of their outcomes for a definite pair of
their local settings. Specifically,~the~latter assumption is abbreviated in
quantum information as the condition on perfect correlations or
anticorrelations.

In this section, we~present the proof~\cite{L3} of the original Bell
inequality in the LHV frame for any numbers of Alice and Bob outcomes in $%
[-1,1]$ and under the condition which is more general than the one
introduced by Bell.

Consider an arbitrary bipartite correlation scenario with two measurement
settings $a_{i},$ $b_{k},$ $i,k=1,2,$ and any numbers of discrete outcomes $%
\lambda _{a},\lambda _{b}\in \lbrack -1,1]$ at Alice and Bob sites,
respectively. This~bipartite scenario is described by four joint
measurements $(a_{i},b_{k}),$ $i,k=1,2,$ with joint probability
distributions $P_{(a_{i},b_{k})}$ of outcomes in $[-1,1]^{2}$. Notation~$%
P_{(a_{i},b_{k})}(\lambda _{a},\lambda _{b})$ means the joint probability of
the event that, under a measurement $(a_{i},b_{k}),$ Alice observes an
outcome $\lambda _{a}$ while Bob---an outcome $\lambda _{b}$. For~the
general framework on the probabilistic description of an arbitrary $N$%
-partite correlation scenario with any numbers of measurement settings and
any spectral type of outcomes at each site, discrete or continuous, see~\cite%
{L31}.

For a joint measurement $(a_{i},b_{k}),$ we denote by%
\begin{equation}
\langle \lambda _{a_{i}}\rangle =\sum_{\lambda _{a},\lambda _{b}\in \lbrack
-1,1]}\lambda _{a}P_{(a_{i},b_{k})}(\lambda _{a},\lambda _{b}),\text{ \ \ \ }%
\langle \lambda _{b_{k}}\rangle =\sum_{\lambda _{a},\lambda _{b}\in \lbrack
-1,1]}\lambda _{b}P_{(a_{i},b_{k})}(\lambda _{a},\lambda _{b})  \label{1_}
\end{equation}%
the averages of outcomes, observed by Alice and Bob, and~by%
\begin{equation}
\langle \lambda _{a_{i}}\lambda _{b_{k}}\rangle =\sum_{\lambda _{a},\lambda
_{b}\in \lbrack -1,1]}\lambda _{a}\lambda _{b}P_{(a_{i},b_{k})}(\lambda
_{a},\lambda _{b})  \label{2_}
\end{equation}%
the average of the product $\lambda _{a}\lambda _{b}$ of their outcomes.

Let, under a joint measurement $(a_{i},b_{k}),$ Alice and Bob outcomes
satisfy the conditions that either the event 
\begin{equation}
\{\lambda _{a}=\lambda _{b}\}:=\left\{ (\lambda _{a},\lambda _{b})\in
\lbrack -1,1]^{2}\mid \lambda _{a}=\lambda _{b}\right\}  \label{3_}
\end{equation}%
or the event 
\begin{equation}
\left\{ \lambda _{a}\text{ }=-\lambda _{b}\neq 0\right\} :=\left\{ (\lambda
_{a},\lambda _{b})\in \lbrack -1,1]^{2}\mid \lambda _{a}=-\lambda _{b}\neq
0\right\}  \label{4_}
\end{equation}%
are observed with certainty, that is~\cite{L3}: 
\begin{eqnarray}
P_{(a_{i},b_{k})}(\{\lambda _{a} &=&\lambda _{b}\})=\sum_{\lambda
_{a}=\lambda _{b}}P_{(a_{i},b_{k})}(\lambda _{a},\lambda _{b})=1  \label{5_}
\\
&&\text{or}  \notag \\
P_{(a_{i},b_{k})}(\{\lambda _{a} &=&-\lambda _{b}\neq 0\})=\sum_{\lambda _{a}%
\text{ }=-\lambda _{b}\neq 0}P_{(a_{i},b_{k})}(\lambda _{a},\lambda _{b})=1,
\label{6_}
\end{eqnarray}%
respectively.

To demonstrate that, under conditions (\ref{5_}) or (\ref{6_}) on
probabilities, outcomes of Alice and Bob are perfectly correlated or
anticorrelated, consider, for~example, the~plus sign case (\ref{5_}). From~(%
\ref{5_}) it follows that, for~arbitrary $\lambda _{a}\neq \lambda _{b},$
the joint probability 
\begin{equation}
P_{(a_{i},b_{k})}(\lambda _{a},\lambda _{b})|_{\lambda _{a}\neq \lambda
_{b}}=0.  \label{7_}
\end{equation}

Hence, under a joint measurement $(a_{i},b_{k}),$ the marginal probabilities
at Alice and Bob sites are given~by%
\begin{eqnarray}
P_{a_{i}}(\lambda _{a}) &=&\sum_{\lambda _{b}}P_{(a_{i},b_{k})}(\lambda
_{a},\lambda _{b})=P_{(a_{i},b_{k})}(\lambda _{a},\lambda _{b})|_{\lambda
_{b}=\lambda _{a}},\text{ \ \ \ }\forall \lambda _{a},  \label{8_} \\
P_{b_{k}}(\lambda _{b}) &=&\sum_{\lambda _{a}}P_{(a_{i},b_{k})}(\lambda
_{a},\lambda _{b})=P_{(a_{i},b_{k})}(\lambda _{a},\lambda _{b})|_{\lambda
_{a}=\lambda _{b}},\text{ \ \ \ }\forall \lambda _{b}.  \notag
\end{eqnarray}

Therefore, under this joint measurement, at~Alice and Bob sites the marginal
probability distributions of observed outcomes $\lambda \in \lbrack -1,1]$
coincide $P_{a_{i}}(\lambda )=P_{b_{k}}(\lambda )$ and, given, for~example,
that Alice observes an outcome $\lambda _{a}=\lambda _{0}$, Bob observes the
outcome $\lambda _{b}=\lambda _{0}$ with certainty, i.e., the~conditional
probability $P_{b_{k}}(\lambda _{b}=\lambda _{0}\mid \lambda _{a}=\lambda
_{0})=1$, $\forall \lambda _{0}.$ Also, under condition (\ref{5_}), the
Pearson correlation coefficient $\gamma _{cor}$, considered in statistics,
is given by 
\begin{equation}
\gamma _{cor}=\frac{\sum_{\lambda _{a},\lambda _{b}}(\lambda _{a}-\langle
\lambda _{a}\rangle )(\lambda _{b}-\langle \lambda _{b}\rangle
)P_{(a_{i},b_{k})}(\lambda _{a},\lambda _{b})}{\sqrt{\sum_{\lambda
_{a}}(\lambda _{a}-\langle \lambda _{a}\rangle )^{2}P_{a_{i}}(\lambda _{a})}%
\sqrt{\sum_{\lambda _{b}}(\lambda _{b}-\langle \lambda _{b}\rangle
)^{2}P_{b_{k}}(\lambda _{b})}}=1.  \label{9_}
\end{equation}

Therefore, under the plus sign condition (\ref{5_}), Alice and Bob outcomes
are perfectly correlated also in the meaning generally accepted in
statistics.

The minus sign case (\ref{6_}) is considered quite similarly and results in
the relation $P_{a_{i}}(\lambda )=P_{b_{k}}(-\lambda ),$ $\forall \lambda
\in \lbrack -1,1],$ for marginal distributions of Alice and Bob, the
relation $P_{b_{k}}(\lambda _{b}=-\lambda _{0}\mid \lambda _{a}=\lambda
_{0})=1$, $\forall \lambda _{0}$, for~the conditional probability and the
Pearson correlation coefficient $\gamma _{cor}=-1.$ All this means the
perfect anticorrelation of Alice and Bob outcomes.

For a joint measurement with outcomes $\pm 1,$ the general conditions (\ref%
{5_}), (\ref{6_}) are equivalently represented by the condition on the
product expectation 
\begin{equation}
\langle \lambda _{a}\lambda _{b}\rangle =\pm 1.  \label{10_}
\end{equation}%
respectively, introduced originally in Bell~\cite{L1}. However,~for~any
number of outcomes in $[-1,1]$ at both sites, Alice and Bob outcomes may be
correlated or anticorrelated in the sense of (\ref{5_}) or (\ref{6_}),
respectively, but~their product expectation $\langle \lambda _{a}\lambda
_{b}\rangle \neq \pm 1.\smallskip $

Thus, under a bipartite scenario with any number of different outcomes in $%
[-1,1]$, relations (\ref{5_}) and (\ref{6_}) introduced in~\cite{L3},
constitute the general condition on perfect correlation or anticorrelation
of outcomes observed by Alice and Bob. This~general perfect
correlations/anticorrelations condition reduces to the Bell one (\ref{10_})
only in a dichotomic case with $\lambda _{a},\lambda _{b}=\pm 1.\smallskip $

Let a $2\times 2$-setting correlation scenario with joint measurements $%
\left( a_{i},b_{k},\right) ,$ $i,k=1,2$ and outcomes $\lambda
_{a_{i}},\lambda _{b_{k}}\in \lbrack -1,1]$ admit a local hidden variable
(LHV) model for joint probabilities, for~details, see~{Section 4} in~\cite%
{L31}, that is, all~joint distributions $P_{(a_{i},b_{k})},i,k=1,2,$ admit
the representation%
\begin{equation}
P_{(a_{i},b_{k})}(\lambda _{a},\lambda _{b})=\int\limits_{\Omega
}P_{a_{i}}(\lambda _{a}|\omega )P_{b_{k}}(\lambda _{b}|\omega )\text{ }\nu (%
\mathrm{d}\omega ),\text{ \ \ }\forall \lambda _{a_{i}},\lambda _{b_{k}},
\label{11_}
\end{equation}%
via a single probability distribution $\nu $ of some variables $\omega \in
\Omega $ and conditional probability distributions $P_{a_{i}}(\mathrm{\cdot }
$ $|\omega ),$ $P_{b_{k}}(\mathrm{\cdot }$ $|\omega )$ of outcomes at
Alice's and Bob's sites. The~latter conditional probabilities are usually
referred to as \textquotedblleft local\textquotedblright\ in the sense that
each of them depends only on a measurement setting at the corresponding site.

Then all scenario product expectations $\langle \lambda _{a_{i}}\lambda
_{b_{k}}\rangle, $ $i,k=1,2,$ admit the LHV representation%
\begin{equation}
\langle \lambda _{a_{i}}\lambda _{b_{k}}\rangle =\int\limits_{\Omega
}f_{a_{i}}(\omega )\text{\ }f_{b_{k}}(\omega )\text{ }\nu (\mathrm{d}\omega )
\label{12_}
\end{equation}%
with 
\begin{equation}
f_{a_{i}}(\omega ):=\sum_{\lambda _{a}\in \lbrack -1,1]}\lambda
_{a}P_{a_{i}}(\lambda _{a}|\omega )\in \lbrack -1,1],\text{ \ \ }%
f_{b_{k}}(\omega ):=\sum_{\lambda _{b}\in \lbrack -1,1]}\lambda
_{b}P_{b_{k}}(\lambda _{b}|\omega )\in \lbrack -1,1].  \label{13_}
\end{equation}

If an LHV model (\ref{11_}) for joint probabilities is {deterministic} ~\cite%
{L4, L31}, then the values of functions $f_{a_{i}},$\ $f_{b_{k}},$ $i,k=1,2,$
constitute outcomes under Alice and Bob corresponding measurements with
settings $a_{i}$ and $b_{k},$ respectively. However,~in~a {stochastic} LHV
model~\cite{L4, L31}, functions $f_{a_{i}},$\ $f_{b_{k}}$ may take any
values in $[-1,1]$ even in a dichotomic case.

On the other side, if, for~a scenario admitting an LHV model (\ref{11_}) and
having outcomes $\lambda _{a_{i}}$, $\lambda _{b_{k}}=\pm 1$, the~Bell
perfect correlation/anticorrelation restriction $\langle \lambda
_{a_{i_{0}}}\lambda _{b_{k_{0}}}\rangle =\pm 1$ is fulfilled under some
joint measurement $(a_{i_{0}}$,$b_{k_{0}}),$ then, in~this LHV\ model,
the~corresponding functions $f_{a_{i_{0}}},f_{b_{k_{0}}}$ take only two
values $\pm 1$ and, moreover, $f_{a_{i_{0}}}(\omega )=\pm
f_{b_{k_{0}}}(\omega ),$ $\nu $-almost everywhere (a.e.) on $\Omega $.

We have the following statement~\cite{L3} (see Appendix, for~the proof).

\begin{Proposition}
\label{p1} Let, under a $2\times 2$-setting correlation scenario with joint
measurements $\left( a_{i},b_{k},\right), $ $i,k=1,2$ and any number of
outcomes $\lambda _{a_{i}},\lambda _{b_{k}}\ $in $[-1,1],$ Alice's and Bob's
outcomes under the joint measurement $(a_{2},b_{1})$ be perfectly correlated
or anticorrelated: 
\begin{eqnarray}
P_{(a_{2},b_{1})}(\{\lambda _{a} &=&\lambda _{b}\})=1  \label{14_1} \\
&&\text{or}  \notag \\
P_{(a_{2},b_{1})}(\{\lambda _{a} &=&-\lambda _{b}\neq 0\})=1  \label{14_3}
\end{eqnarray}

If this scenario admits an LHV model (\ref{11_}), then its product
expectations satisfy the original Bell~inequality: 
\begin{equation}
\left\vert \text{ }\langle \lambda _{a_{1}}\lambda _{b_{1}}\rangle -\langle
\lambda _{a_{1}}\lambda _{b_{2}}\rangle \right\vert \pm \langle \lambda
_{a_{2}}\lambda _{b_{2}}\rangle \leq 1,  \label{15_}
\end{equation}%
in its perfect correlation (plus sign) or perfect anticorrelation (minus
sign) forms, respectively.
\end{Proposition}

We stress that, for~the validity of the original Bell inequality (\ref{15_})
in the LHV frame, it~is suffice for condition (\ref{14_1}) or condition (\ref%
{14_3}) on perfect correlations or anticorrelations be fulfilled only under
a joint measurement $(a_{2},b_{1})$.

Furthermore, it~was proved in~\cite{L3} that, in~the LHV frame, the~original
Bell inequality~(\ref{15_}) holds under the LHV condition which is more
general than conditions~(\ref{14_1}), (\ref{14_3}) on perfect
correlation/anticorrelations, does not imply for the LHV functions (\ref{13_}%
) relations $f_{a_{2}}(\omega )=\pm f_{b_{1}}(\omega ),$ $\nu $-$a.e.$ on $%
\Omega $ and incorporates conditions (\ref{14_1}), (\ref{14_3}) on perfect
correlation/anticorrelations only as particular cases.

For many bipartite quantum states admitting $2\times 2$-setting LHV models,
specifically, this general sufficient condition in~\cite{L3} ensures~\cite%
{L51, L3, L5} the validity of the perfect correlation form of the original
Bell inequality for Alice and Bob measurements for any three qudit quantum
observables \mbox{$X_{a_{1}},X_{a_{2}}=X_{b_{1}},X_{b_{2}}$} with operator
norms $\leq 1$. Satisfying~the perfect correlation form of the original Bell
inequality (\ref{15_}), these states do not need to exhibit perfect
correlations and may even have a negative correlation function (see relation
(61) in~\cite{L3}) whenever the same quantum observable $X_{a_{2}}=X_{b_{1}}$
is measured at both sites.

For example, all~two-qudit Werner state~\cite{L6}%
\begin{equation}
W_{d,\Phi }=\frac{1+\Phi }{2}\frac{\mathrm{P}_{d}^{(+)}}{r_{d}^{(+)}}+\frac{%
1-\Phi }{2}\frac{\mathrm{P}_{d}^{(-)}}{r_{d}^{(-)}},\text{ \ \ \ }\Phi \in
\lbrack -1,1],  \label{19_}
\end{equation}%
on $\mathbb{C}^{d}\otimes \mathbb{C}^{d},$ $d\geq 3,$ separable ($\Phi \in
\lbrack 0,1]$) or nonseparable ($\Phi \in \lbrack -1,0)$), and~all separable
two-qubit Werner stated $W_{2,\Phi }(\Phi ),$ $\Phi \in \lbrack 0,1],$
satisfy the general sufficient condition, introduced in~\cite{L3}, and~do
not violate the perfect correlation form of the original Bell inequality (%
\ref{15_}) for any three quantum observables $%
X_{a_{1}},X_{a_{2}}=X_{b_{1}},X_{b_{2}}$ but do not exhibit perfect
correlations whenever the same observable $X_{a_{2}}=X_{b_{1}}$ is measured
at both sites. \ In (\ref{19_}), $\mathrm{P}_{d}^{(\pm )}$ are the
orthogonal projections onto the symmetric and antisymmetric subspaces of $%
\mathbb{C}^{d}\otimes \mathbb{C}^{d}$ with dimensions $r_{d}^{(\pm )}=%
\mathrm{tr}[\mathrm{P}_{d}^{(\pm )}]$ $=\frac{d(d\pm 1)}{2},$ respectively.

\section{Quantum Violation \label{s3}}

Consider Alice and Bob projective measurements of quantum qudit observable $%
X_{a_{1}},X_{a_{2}}=X_{b_{1}},$ $X_{b_{2}}$ in an arbitrary two-qudit state $%
\rho $ on $\mathbb{C}^{d}\otimes \mathbb{C}^{d}$.

In this case, Alice and Bob outcomes coincide with eigenvalues $\lambda
_{a},\lambda _{b}$ of these observables and restriction $\lambda
_{a},\lambda _{b}\in \lbrack -1,1]$ implies the restriction on operators
norms $\left\Vert X_{a_{i}}\right\Vert, \left\Vert X_{b_{k}}\right\Vert \leq
1$. The~joint probability $P_{(a_{i},b_{k})}(\lambda _{a},\lambda _{b})$
that, under a joint measurement $(a_{i},b_{k}),$ Alice observes an outcome $%
\lambda _{a},$ while Bob---and outcome $\lambda _{b}$ is given by 
\begin{equation}
\mathrm{tr}[\rho \{\mathrm{P}_{X_{a_{i}}}(\lambda _{a})\otimes \mathrm{P}%
_{X_{b_{k}}}(\lambda _{b})\}]  \label{20_1}
\end{equation}%
where $\mathrm{P}_{X_{a_{i}}}(\lambda _{a}),$ $\mathrm{P}_{X_{b_{k}}}(%
\lambda _{b})$, $i,k=1,2,$ are the spectral projections of observables $%
X_{a_{i}}$ and $X_{b_{k}},$ corresponding to eigenvalues $\lambda _{a}$ and $%
\lambda _{b}$, respectively. The~averages in (\ref{1_}), (\ref{2_}) take the
form \ 
\begin{equation}
\langle \lambda _{a_{i}}\rangle =\mathrm{tr}[\rho X_{a_{i}}],\text{ \ \ \ }%
\langle \lambda _{b_{k}}\rangle =\mathrm{tr}[\rho X_{b_{k}}],\text{ \ \ \ \ }%
\langle \lambda _{a_{i}}\lambda _{b_{k}}\rangle =\mathrm{tr}[\rho
\{X_{a_{i}}\otimes X_{b_{k}}\}],\text{ }i,k=1,2  \label{20_}
\end{equation}

The general conditions (\ref{14_1}), (\ref{14_3}) on perfect correlations or
anticorrelations of Alice and Bob outcomes under a joint measurement $%
(a_{2},b_{1})$ reduce to 
\begin{eqnarray}
\sum_{\lambda _{a}=\lambda _{b}}\mathrm{tr}[\rho \{\mathrm{P}%
_{X_{b_{1}}}(\lambda _{a})\otimes \mathrm{P}_{X_{b_{1}}}(\lambda _{b})\}]
&=&1,  \label{22_1} \\
\sum_{\lambda _{a}=-\lambda _{b}\neq 0}\mathrm{tr}[\rho \{\mathrm{P}%
_{X_{b_{1}}}(\lambda _{a})\otimes \mathrm{P}_{X_{b_{1}}}(\lambda _{b})\}]
&=&1,  \label{22_2}
\end{eqnarray}%
respectively, and~for observables with eigenvalues $\pm 1,$ these conditions
are equivalent to%
\begin{equation}
\mathrm{tr}[\rho \{X_{b_{1}}\otimes X_{b_{1}}\}]=\pm 1.  \label{22_}
\end{equation}

Thus, under the considered quantum scenario, the~left hand-side $W_{\rho
_{d}}^{(\pm )}$ of the original Bell inequality (\ref{15_}) takes the form 
\begin{equation}
W_{\rho }^{(\pm )}(X_{a},X_{b_{1}},X_{b_{2}})=\left\vert \text{ }\mathrm{tr}%
[\rho \{X_{a}\otimes X_{b_{1}}\}]-\mathrm{tr}[\rho \{X_{a}\otimes
X_{b_{2}}\}]\text{ }\right\vert \pm \mathrm{tr}[\rho \{X_{b_{1}}\otimes
X_{b_{2}}\}],  \label{22_3}
\end{equation}%
where, for~short, we~changed the index notation $a_{1}\rightarrow a,$and the
general condition on perfect correlations/anticorrelations of Alice and Bob
outcomes under a joint measurement $(b_{1},b_{1})$ is given by~(\ref{22_1})/(%
\ref{22_2}).

It is, however, well known that the two-qubit singlet state $\rho _{singlet}$
satisfies the perfect anticorrelation (minus sign) condition (in the form (%
\ref{22_})) whenever the same qubit observable $X_{b}$ with eigenvalues $\pm
1$ is measured at both sites but, depending on a choice of qubit observables 
$X_{a},X_{b_{1}},X_{b_{2}}$, this state may, however, violate~\cite{L1, L2}
the perfect anticorrelation form of the original Bell inequality (\ref{15_}).

As it has been proven in~\cite{L7, L8}, for~the singlet $\rho _{singlet}$,
the maximal value of the left hand-side~(\ref{22_3}) of the original Bell
inequality (\ref{15_}) over qubit observables with eigenvalues $\pm 1$ is
equal to $\frac{3}{2}$.

This value is beyond the well-known Tsirelson~\cite{L9, L10} maximal value $%
\sqrt{2}$ for the quantum violation parameter $\left\vert \mathcal{B}%
_{quant}^{\mathrm{CHSH}}\right\vert /\left\vert \mathcal{B}_{lhv}^{\mathrm{%
CHSH}}\right\vert $ of the Clauser--Horne--Shimony--Holt (CHSH) inequality~%
\cite{CHSH} $\left\vert \mathcal{B}_{lhv}^{CHSH}\right\vert \leq 2$ and,
moreover, beyond the least upper bound $\sqrt{2}$ on the quantum violation
parameter $\left\vert \mathcal{B}_{quant}\right\vert /\left\vert \mathcal{B}%
_{lhv}\right\vert $ for all unconditional Bell functionals $\mathcal{B(\cdot
)}$ for two settings and two {outcomes per site}~\cite{L12, L13, L14, L15}. 
%

On the other side, the~Tsirelson bound $2\sqrt{2}$ on the quantum violation
of the CHSH inequality~\cite{CHSH} holds for a bipartite quantum state of an
arbitrary dimension. For~different choices of signs, this~implies%
\begin{equation}
\begin{array}{ccc}
\mathrm{tr}[\rho \{X_{a}\otimes X_{b_{1}}\}]-\mathrm{tr}[\rho \{X_{a}\otimes
X_{b_{2}}\}+\mathrm{tr}[\rho \{X_{b_{1}}\otimes X_{b_{1}}\}+\mathrm{tr}[\rho
\{X_{b_{1}}\otimes X_{b_{2}}\}] & \leq & 2\sqrt{2} \label{23_} \\ 
\mathrm{tr}[\rho \{X_{a}\otimes X_{b_{1}}\}]-\mathrm{tr}[\rho \{X_{a}\otimes
X_{b_{2}}\}-\mathrm{tr}[\rho \{X_{b_{1}}\otimes X_{b_{1}}\}-\mathrm{tr}[\rho
\{X_{b_{1}}\otimes X_{b_{2}}\}] & \leq & 2\sqrt{2} \\ 
-\mathrm{tr}[\rho \{X_{a}\otimes X_{b_{1}}\}]+\mathrm{tr}[\rho
\{X_{a}\otimes X_{b_{2}}\}+\mathrm{tr}[\rho \{X_{b_{1}}\otimes X_{b_{1}}\}+%
\mathrm{tr}[\rho \{X_{b_{1}}\otimes X_{b_{2}}\}] & \leq & 2\sqrt{2} \\ 
-\mathrm{tr}[\rho \{X_{a}\otimes X_{b_{1}}\}]+\mathrm{tr}[\rho
\{X_{a}\otimes X_{b_{2}}\}-\mathrm{tr}[\rho \{X_{b_{1}}\otimes X_{b_{1}}\}-%
\mathrm{tr}[\rho \{X_{b_{1}}\otimes X_{b_{2}}\}] & \leq & 2\sqrt{2}%
\end{array}%
\end{equation}

Combining the first line with the third one, for~a two-qudit state
exhibiting perfect correlations (condition (\ref{22_1})), we~get the
following upper bound 
\begin{equation}
\begin{array}{ccl}
\vspace{6pt} W_{\rho }^{(+)}(X_{a},X_{b_{1}},X_{b_{2}})|_{perfect} & = & 
\left\vert \text{ }\mathrm{tr}[\rho \{X_{a}\otimes X_{b_{1}}\}]-\mathrm{tr}%
[\rho \{X_{a}\otimes X_{b_{2}}\}\right\vert +\mathrm{tr}[\rho
\{X_{b_{1}}\otimes X_{b_{2}}\}] \label{24_1} \\ 
& \leq & 2\sqrt{2}-\left\vert \text{ }\mathrm{tr}[\rho \{X_{b_{1}}\otimes
X_{b_{1}}\}]\right\vert%
\end{array}%
\end{equation}
on the left-hand side of the original Bell inequality. Similarly,~combining
the second line with the fourth one\ under condition (\ref{22_2}) on perfect
anticorrelations, we~derive%
\begin{equation}
\begin{array}{ccl}
\vspace{6pt} W_{\rho }^{(-)}(X_{a},X_{b_{1}},X_{b_{2}})|_{perfect} & = & 
\left\vert \text{ }\mathrm{tr}[\rho \{X_{a}\otimes X_{b_{1}}\}]-\mathrm{tr}%
[\rho \{X_{a}\otimes X_{b_{2}}\}\right\vert -\mathrm{tr}[\rho
\{X_{b_{1}}\otimes X_{b_{2}}\}] \label{24_2} \\ 
& \leq & 2\sqrt{2}-\left\vert \text{ }\mathrm{tr}[\rho \{X_{b_{1}}\otimes
X_{b_{1}}\}]\right\vert%
\end{array}%
\end{equation}

Thus, for~an arbitrary two-qudit state exhibiting perfect
correlation/anticorrelations whenever the same quantum observable $X_{b_{1}}$
is measured at both sites we have%
\begin{equation}
\begin{array}{ccl}
\vspace{6pt} W_{\rho }^{(\pm )}(X_{a},X_{b_{1}},X_{b_{2}})|_{perfect} & = & 
\left\vert \text{ }\mathrm{tr}[\rho \{X_{a}\otimes X_{b_{1}}\}]-\mathrm{tr}%
[\rho \{X_{a}\otimes X_{b_{2}}\}]\text{ }\right\vert \pm \mathrm{tr}[\rho
\{X_{b_{1}}\otimes X_{b_{2}}\}] \label{24_3} \\ 
& \leq & 2\sqrt{2}-\left\vert \text{ }\mathrm{tr}[\rho \{X_{b_{1}}\otimes
X_{b_{1}}\}]\right\vert%
\end{array}%
\end{equation}

If observable $X_{b_{1}}$ has only eigenvalues $\pm 1,$ then conditions (\ref%
{22_1}), (\ref{22_2}) reduce to the Bell condition~(\ref{22_}) and the upper
bound (\ref{24_3}) takes the form%
\begin{equation}
W_{\rho }^{(\pm )}(X_{a},X_{b_{1}},X_{b_{2}})|_{perfect}\leq 2\sqrt{2}-1
\label{25_}
\end{equation}%
and holds for a two-qudit state $\rho $ of an arbitrary dimension $d\geq 2$.
For $d=2$, this upper bound is more than the maximal value $\frac{3}{2}$
proved~\cite{L7, L8} for the two-qubit singlet.

Therefore, in~the following section, we~proceed to analyze the maximal value
which the left-hand of $W_{\rho }^{(\pm )}(X_{a},X_{b_{1}},X_{b_{2}})|_{%
\text{\emph{perfect}}}$ over all qubit observables $%
X_{a},X_{b_{1}},X_{b_{2}} $ with eigenvalues $\pm 1$ and all two-qubit
states $\rho, $ satisfying the perfect correlation/anticorrelation condition
(\ref{22_}).

\section{Two-Qubit Case \label{s4}}

Consider the violation of the original Bell inequality (\ref{15_}) by a
two-qubit state exhibiting perfect correlations/anticorrelations whenever
the same qubit quantum observable with eigenvalues $\pm 1$ is projectively
measured at both sites.

We further consider only symmetric two-qubit states $\rho $ (identical
quantum particles), \ that is, states on $\mathbb{C}^{2}\otimes \mathbb{C}%
^{2}$ which do not change under the permutation of the Hilbert spaces $%
\mathbb{C}^{2}$ in the tensor product $\mathbb{C}^{2}\otimes \mathbb{C}^{2},$
and, for~simplicity, change index notations $b_{1}\rightarrow
r,b_{2}\rightarrow c$ in (\ref{22_3}).

For $d=2$, a~generic qubit observable $X$ on $\mathbb{C}^{2}$ admits the
representation 
\begin{eqnarray}
X &=&\alpha \mathbb{I}_{\mathbb{C}^{2}}+r\cdot \sigma ,\text{ \ \ \ }
\label{26_} \\
r\cdot \sigma &=&r_{1}\sigma _{1}+r_{2}\sigma _{2}+r_{3}\sigma _{3}
\label{26_1}
\end{eqnarray}%
where $\alpha =\frac{1}{2}\mathrm{tr}[X],$ $r=(r_{1},r_{2},r_{3})$ is a
vector in $\mathbb{R}^{3}$ with components 
\begin{equation}
r_{1}=\frac{1}{2}\mathrm{tr}[X\sigma _{1}],\text{ \ \ }r_{2}=\frac{1}{2}%
\mathrm{tr}[X\sigma _{2}],\text{ \ \ }r_{3}=\frac{1}{2}\mathrm{tr}[X\sigma
_{3}],  \label{27_}
\end{equation}%
and%
\begin{equation}
\sigma _{1}=|e_{1}\rangle \langle e_{2}|\text{ }+\text{ }|e_{2}\rangle
\langle e_{1}|,\text{ \ \ \ }\sigma _{2}=i(|e_{2}\rangle \langle e_{1}|\text{
}-|e_{1}\rangle \langle e_{2}|),\text{ \ \ }\sigma _{3}=|e_{1}\rangle
\langle e_{1}|\text{ }-\text{ }|e_{2}\rangle \langle e_{2}|  \label{28_1}
\end{equation}%
are self-adjoint operators on\ $\mathbb{C}^{2}$ with eigenvalues $\pm 1,$
represented in the standard orthonormal basis $\{e_{1},e_{2}\}$ in $\mathbb{C%
}^{2}$ by the Pauli matrices 
\begin{equation}
\sigma _{1}=%
\begin{pmatrix}
0 & 1 \\ 
1 & 0%
\end{pmatrix}%
,\ \sigma _{2}=%
\begin{pmatrix}
0 & -i \\ 
i & 0%
\end{pmatrix}%
,\ \sigma _{3}=%
\begin{pmatrix}
1 & 0 \\ 
0 & -1%
\end{pmatrix}%
.  \label{28_2}
\end{equation}

Every qubit observable with eigenvalues $\pm 1$ is represented in (\ref{26_}%
) by some unit vector $\left\Vert r\right\Vert =1$ and constitutes
projection $\sigma _{r}:=r\cdot \sigma $ of the qubit spin along a unit
vector (direction) $r$ in $\mathbb{R}^{3}.$

Therefore, for~Alice and Bob measurements of qubit observables with
eigenvalues $\pm 1,$ the left-hand side (\ref{22_3}) of the original Bell
inequality takes the form%
\begin{equation}
W_{\rho }^{(\pm )}(\sigma _{a},\sigma _{r},\sigma _{c})=\left\vert \text{ }%
\mathrm{tr}[\rho \{\sigma _{a}\otimes \sigma _{r}\}]-\mathrm{tr}[\rho
\{\sigma _{a}\otimes \sigma _{c}\}]\text{ }\right\vert \pm \mathrm{tr}[\rho
\{\sigma _{r}\otimes \sigma _{c}\}]  \label{29_}
\end{equation}%
where $a,r,c$ are unit vectors in $\mathbb{R}^{3}$ and the relation 
\begin{equation}
\mathrm{tr}[\rho \{\sigma _{r}\otimes \sigma _{r}\}]=\pm 1  \label{30_}
\end{equation}%
constitutes the perfect correlation/anticorrelation of Alice and Bob
outcomes whenever the same spin observable $\sigma _{r}$---the projection of
qubit spin along the same direction $r$ in $\mathbb{R}^{3}$---is measured at
both sites.

Substituting representation (\ref{26_1}) into (\ref{29_}) and (\ref{30_}),
we rewrite these relations via scalar products of vectors in $\mathbb{R}%
^{3}: $ 
\begin{eqnarray}
W_{\rho }^{(\pm )}(\sigma _{a},\sigma _{r},\sigma _{c}) &=&\left\vert
(a,T^{(\rho )}r)-(a,T^{(\rho )}c)\right\vert \pm (r,T^{(\rho )}c),
\label{31_} \\
(r,T^{(\rho )}r) &=&\pm 1,  \label{32_}
\end{eqnarray}%
where $(a,T^{(\rho )}r):=\sum_{i,j}T_{ij}^{(\rho )}a_{i}r_{j}$ and $T^{(\rho
)}$ is the linear operator on $\mathbb{R}^{3},$ defined in the canonical
basis in $\mathbb{R}^{3}$ by the matrix with real elements 
\begin{equation}
T_{ij}^{(\rho )}:=\mathrm{tr}[\rho \{\sigma _{i}\otimes \sigma _{j}\},\text{
\ }i,j=1,2,3,  \label{33_}
\end{equation}

This correlation matrix is symmetric (since $\rho $\ is symmetric), has
eigenvalues $\lambda _{m},$\ $m=1,2,3,$\ where all $\left\vert \lambda
_{m}\right\vert \leq 1,$\ and is similar by its form to the matrix
considered in~\cite{L16}.

Let us first analyze when an arbitrary symmetric two-qubit state $\rho $ may
satisfy condition (\ref{32_}). By~decomposing a unit vector $r=\sum_{m}\beta
_{m}\mathrm{v}_{m}$, $\sum_{m}\beta _{m}^{2}=1,$ in the orthonormal basis $\{%
\mathrm{v}_{j},j=1,2,3\}$ of eigenvectors of $T^{(\rho )},$ we rewrite
condition (\ref{32_}) in the form%
\begin{equation}
\sum_{m}\beta _{m}^{2}(\lambda _{m}\mp 1)=0.  \label{35_}
\end{equation}

Since all eigenvalues $\left\vert \lambda _{m}\right\vert \leq 1$, relation (%
\ref{35_}) implies the following statement.

\begin{Proposition}
\label{p2} A symmetric two-qubit state $\rho $ exhibits perfect
correlation/anticorrelations 
\begin{equation}
\mathrm{tr}[\rho \{\sigma _{r}\otimes \sigma _{r}\}]=\pm 1  \label{36_}
\end{equation}%
\vspace{6pt} if and only if its correlation matrix $T^{(\rho )}$ has at
least one eigenvalue equal to $\pm 1,$ respectively. In~this case:

$\mathrm{(1)}$ if only one of eigenvalues of $T^{(\rho )}$ is equal to $\pm
1, $ say $\lambda _{m_{0}}=\pm 1,$ then $\rho $ satisfies the perfect
correlation/anticorrelation condition (\ref{36_}), respectively, only for
the unit vector $r=\mathrm{v}_{m_{0}}$;

$\mathrm{(2)}$ if $T^{(\rho )}$ has two eigenvalues equal to $\pm 1,$ say $%
\lambda _{m_{1}},\lambda _{m_{2}}=$ $\pm 1,$ then $\rho $ satisfies the
perfect correlation/anticorrelation condition (\ref{36_}), respectively for
every unit vector $r=\beta _{m_{1}}\mathrm{v}_{m_{1}}+\beta _{m_{2}}\mathrm{v%
}_{m_{2}},$ \mbox{$\beta _{m_{1}}^{2}+\beta _{m_{2}}^{2}=1$} in the plane
determined by the eigenvectors $\{\mathrm{v}_{m_{1}},\mathrm{v}_{m_{2}}\}$
of $T^{(\rho )}$;

$\mathrm{(3)}$ if all three eigenvalues of $T^{(\rho )}$ are equal to $\pm
1, $ then $\rho $ satisfies the perfect correlation/anticorrelation
condition (\ref{36_}), respectively, for~any unit vector $r$ in $\mathbb{R}%
^{3}.$
\end{Proposition}

For the two-qubit Bell states 
\begin{equation}
\phi _{(\pm )}=\frac{1}{\sqrt{2}}\left( e_{1}\otimes e_{1}\pm e_{2}\otimes
e_{2}\right) \text{, \ \ \ }\psi _{(\pm )}=\frac{1}{\sqrt{2}}\left(
e_{1}\otimes e_{2}\pm e_{2}\otimes e_{1}\right),  \label{37_}
\end{equation}%
we have%
\begin{equation}
\begin{array}{ccc}
\vspace{6pt} T^{(\phi _{+})} & = & 
\begin{pmatrix}
1 & 0 & 0 \\ 
0 & -1 & 0 \\ 
0 & 0 & 1%
\end{pmatrix}%
,\text{ \ \ \ }T^{(\phi _{-})}=%
\begin{pmatrix}
-1 & 0 & 0 \\ 
0 & 1 & 0 \\ 
0 & 0 & 1%
\end{pmatrix}
\label{38_} \\ 
T^{(\psi _{+})} & = & 
\begin{pmatrix}
1 & 0 & 0 \\ 
0 & 1 & 0 \\ 
0 & 0 & -1%
\end{pmatrix}%
,\text{ \ \ \ }T^{(\psi _{-})}=%
\begin{pmatrix}
-1 & 0 & 0 \\ 
0 & -1 & 0 \\ 
0 & 0 & -1%
\end{pmatrix}%
\end{array}%
\end{equation}

and this implies.

\begin{Corollary}
(1) The Bell state $\phi _{+}$ exhibits perfect anticorrelations under spin
measurements at both sites along the coordinate axis Y and perfect
correlations under spin measurements at both sites along the same arbitrary
direction in the coordinate plane \textrm{XZ};

(2) The Bell state $\phi _{-}$ exhibits perfect anticorrelations under spin
measurements at both sites along the coordinate axis \textrm{X} and perfect
correlations---under spin measurements at both sites along the same
arbitrary direction in the coordinate plane \textrm{YZ};

(3) The Bell state $\psi _{+}$ exhibits perfect anticorrelations under
measurements at both sites of spin projections along the coordinate axis 
\textrm{Z} and perfect correlations---under spin measurements at both along
the same arbitrary direction in the coordinate plane \textrm{XY};

(4) The Bell state (singlet) $\psi _{-}$ exhibits perfect anticorrelations
under spin measurements at both sites along the same arbitrary direction in $%
\mathbb{R}^{3}.$
\end{Corollary}

Let us now analyze the maximal value of the left-hand side (\ref{31_}) of
the original Bell inequality for a two-qubit state $\rho $ exhibiting
perfect correlations/anticorrelations (\ref{32_}).

Under condition $\left\Vert a\right\Vert =1,$ the maximum of $W_{\rho
}^{(\pm )}(\sigma _{a},\sigma _{r},\sigma _{c})$ over $a$ is reached on the
unit vector 
\begin{equation}
a=\pm \frac{T^{(\rho )}(r-c)}{\left\Vert T^{(\rho )}(r-c)\right\Vert }
\label{41_}
\end{equation}%
and is given by%
\begin{equation}
\left\Vert T^{(\rho )}(r-c)\right\Vert \pm (r,T^{(\rho )}c).  \label{42_}
\end{equation}

Expanding vectors $r=\sum_{m}\beta _{m}\mathrm{v}_{m}$, $\sum \beta
_{m}^{2}=1,$ $c=\sum_{m}\gamma _{m}\mathrm{v}_{m}$, $\sum_{m}\gamma
_{m}^{2}=1,$ in terms of the orthonormal eigenvectors $\left\{ \mathrm{v}%
_{m}\right\} $ of $T^{(\rho )},$ we rewrite (\ref{42_}) in the form%
\begin{equation}
\sqrt{\sum_{m=1,2,3}\lambda _{m}^{2}(\beta _{m}-\gamma _{m})^{2}}\pm
\sum_{m=1,2,3}\lambda _{m}\beta _{m}\gamma _{m},  \label{43_}
\end{equation}%
where, due~to perfect correlations/anticorrelations condition (\ref{32_}),
the coefficients $\beta _{m}$ are specified in {Proposition \ref{p2}}.

Consider the maximum of expression (\ref{43_}) over coefficients $\gamma
_{m}.$ By Proposition \ref{p2}, expression (\ref{43_}) reduces to 
\begin{equation}
\begin{array}{ccc}
\vspace{9pt} &  & \sqrt{\sum_{\lambda _{m}^{2}=1}(\beta _{m}-\gamma
_{m})^{2}+\sum_{\lambda _{m}^{2}\neq 1}\lambda _{m}^{2}\gamma _{m}{}^{2}}%
+\sum_{\lambda _{m}^{2}=1}\beta _{m}\gamma _{m} \label{44_} \\ 
& = & \sqrt{2(1-\sum_{\lambda _{m}^{2}=1}\beta _{m}\gamma
_{m})-\sum_{\lambda _{m}^{2}\neq 1}(1-\lambda _{m}^{2})\gamma _{m}^{2}}%
+\sum_{\lambda _{m}^{2}=1}\beta _{m}\gamma _{m}%
\end{array}%
\end{equation}
since $\sum_{\lambda _{m}^{2}=1}\beta _{m}^{2}=1.$ From (\ref{44_}) it
follows that, for~all choices of a direction $r$---coefficients $\beta _{m}$
in (\ref{44_}) specified in {Proposition \ref{p2}}, we~have

\begin{equation}
\sup_{a,c}W_{\rho }^{(\pm )}(\sigma _{a},\sigma _{r},\sigma _{c})|_{\text{%
\emph{perfect}}}\leq \max_{z\in \lbrack -1,1]}\left( \sqrt{2(1-z)}+z\right) =%
\frac{3}{2}  \label{46_}
\end{equation}%
where the upper bound $\frac{3}{2}$ is, for~example, reached on every Bell
state where all eigenvalues of the correlation matrices {$\lambda _{m}\in
\{-1,1\},m=1,2,3.$}

Also, if~a two-qubit state, exhibiting perfect correlations/anticorrelations
(see Proposition \ref{p2}), has~the correlation matrix with at least two
eigenvalues, say~$\lambda _{m_{1}},\lambda _{m_{2}},$ with $\left\vert
\lambda _{m_{1}}\right\vert ,\left\vert \lambda _{m_{2}}\right\vert =1,$
then the upper bound $\frac{3}{2}$ is reached on the unit vector $c$ which
is in the plane of eigenvectors $\mathrm{v}_{m_{1}},\mathrm{v}_{m_{2}}$
corresponding to these eigenvalues (vector $r$ is in this plane, see
Proposition 2) and satisfies condition $c\cdot r=$ $\sum_{\lambda
_{m}^{2}=1}\beta _{m}\gamma _{m}=\frac{1}{2},$ that is, at~angle $\pi /3$ to
vector $r.$

Thus, we~have proved the following new result.

\begin{Theorem}
\label{theorem 1} Let $\rho \mathfrak{\ }$be a symmetric two-qubit states on 
$\mathbb{C}^{2}\otimes \mathbb{C}^{2}$ exhibiting perfect
correlations/anticorrelations whenever the same qubit observable $\sigma
_{r} $ is measured at both sites. Then the maximal value of the left-hand
side $W_{\rho }^{(\pm )}(\sigma _{a},\sigma _{r},\sigma _{c})$ of the
original Bell inequality is given by 
\begin{equation}
\max_{\rho, a,r,c}W_{\rho }^{(\pm )}(\sigma _{a},\sigma _{r},\sigma _{c})|_{%
\text{perfect}}=\frac{3}{2}  \label{47_}
\end{equation}%
and is reached on symmetric two-qubit states discussed in lines after
Equation (\ref{46_}).
\end{Theorem}

We stress that this maximal value is less than the upper bound (\ref{25_})
following from the CHSH~inequality.

\section{Two-Qutrit Case \label{s5}}

Consider now the violation of the original Bell inequality under Alice and
Bob spin measurements on a symmetric two-qutrit state $\rho $ on $\mathbb{C}%
^{3}\otimes \mathbb{C}^{3},$ exhibiting perfect correlations or
anticorrelations.

For Alice and Bob spin measurements in a two-qutrit state $\rho $, the
left-hand side (\ref{22_3}) of the original Bell inequality and the
condition on perfect correlations/anticorrelations take the forms 
\begin{eqnarray}
W_{\rho }^{(\pm )}(S_{a},S_{r},S_{c}) &=&\left\vert \text{ }\mathrm{tr}[\rho
\{S_{a}\otimes S_{r}\}]-\mathrm{tr}[\rho \{S_{a}\otimes S_{c}\}]\text{ }%
\right\vert \pm \mathrm{tr}[\rho \{S_{r}\otimes S_{c}\}],  \label{48_1} \\
\mathrm{tr}[\rho \{S_{r}\otimes S_{r}\}] &=&\pm 1,  \label{48_2}
\end{eqnarray}%
where $a,r,c$ are unit vectors in $\mathbb{R}^{3}$ and 
\begin{equation}
S_{r}=r\cdot S=r_{1}S_{1}+r_{2}S_{2}+r_{3}S_{3},\text{ \ }\
S=(S_{1},S_{2},S_{3}),  \label{49_}
\end{equation}%
is the qutrit observable with eigenvalues $\{1,0,-1\},$ describing
projection of qutrit spin along a unit vector $r$ in $\mathbb{R}^{3}$.

Note that if a two-qutrit state $\rho $ exhibits perfect
correlations/anticorrelations (\ref{48_2}) under measurements in this state
at both sites of spin projection along a direction $r$, the~probability of
event that either Alice or Bob observe at their site the outcome $\lambda =0$
is equal to zero.

In the standard orthonormal basis $\{e_{1},e_{2},e_{3}\}$ in $\mathbb{C}^{3}$
these operators have the following matrix~representations: 
\begin{equation}
S_{1}=\frac{1}{\sqrt{2}}%
\begin{pmatrix}
0 & 1 & 0 \\ 
1 & 0 & 1 \\ 
0 & 1 & 0%
\end{pmatrix}%
,\text{ \ \ }S_{2}=\frac{1}{\sqrt{2}}%
\begin{pmatrix}
0 & -i & 0 \\ 
i & 0 & -i \\ 
0 & i & 0%
\end{pmatrix}%
,\text{ \ \ }S_{3}=%
\begin{pmatrix}
1 & 0 & 0 \\ 
0 & 0 & 0 \\ 
0 & 0 & -1%
\end{pmatrix}
\label{50_}
\end{equation}%
and 
\begin{equation}
\text{\ }S_{r}=%
\begin{pmatrix}
r_{3} & \frac{r_{1}-ir_{2}}{\sqrt{2}} & 0 \\ 
\frac{r_{1}+ir_{2}}{\sqrt{2}} & 0 & \frac{r_{1}-ir_{2}}{\sqrt{2}} \\ 
0 & \frac{r_{1}+ir_{2}}{\sqrt{2}} & -r_{3}%
\end{pmatrix}
\label{51_}
\end{equation}

In view of (\ref{49_}), quite similarly to our techniques in Section~\ref{s4}
we introduce for a symmetric two-qutrit state $\rho $ the correlation matrix 
$Z^{(\rho )}$ with real elements 
\begin{equation}
Z_{ij}^{(\rho )}=\mathrm{tr}[\rho \{S_{i}\otimes S_{j}\}],  \label{52_}
\end{equation}%
which is symmetric, diagonalized and has eigenvalues $\left\vert \lambda
_{m}\right\vert \leq 1,$ and this allows us to rewrite (\ref{48_1}), (\ref%
{48_2}) in the form:

\begin{equation}
\begin{array}{rll}
\vspace{6pt} W_{\rho }^{(\pm )}(S_{a},S_{r},S_{c}) & = & \left\vert
(a,Z^{(\rho )}r)-(a,Z^{(\rho )}c)\right\vert \pm (r,Z^{(\rho )}c), \label%
{53_} \\ 
(r,Z^{(\rho )}r) & = & \pm 1.%
\end{array}%
\end{equation}
These expressions are quite the same by their form to expressions (\ref{31_}%
), (\ref{32_}) for a two-qubit state. By using the same techniques as in a
qubit case, we~derive 
\begin{equation}
\sup_{a,c}W_{\rho }^{(\pm )}(S_{a},S_{r},S_{c})|_{\text{\emph{perfect}}}\leq 
\frac{3}{2}.  \label{54_2}
\end{equation}%
We, however, do~not know whether under the considered measurements this
supremum is~reached.

\begin{Theorem}
Let $\rho \mathfrak{\ }$be a symmetric two-qutrit states on $\mathbb{C}%
^{3}\otimes \mathbb{C}^{3}$ exhibiting perfect correlations/anticorrelations
whenever spin projection $S_{r}$ along a direction $r$ is measured at both
sites. Then,~under Alice and Bob spin measurements on these two-qutrit
states, the~maximal value of the left-hand side $W_{\rho }^{(\pm
)}(S_{a},S_{r},S_{c})$ of the original Bell inequality (\ref{15_}) is upper
bounded as 
\begin{equation}
\sup_{\rho, a,r,c}W_{\rho }^{(\pm )}(S_{a},S_{r},S_{c})|_{\text{perfectBell}%
}\leq \frac{3}{2}.  \label{54_1}
\end{equation}
\end{Theorem}

This two-qutrit upper bound is less than the upper bound (\ref{25_})
following from the CHSH~inequality.

\section{Conclusions \label{s6}}

As was pointed out in the Introduction, the~recent tremendous developments
in quantum technologies make experiments to test the original Bell
inequality at least less difficult. This~stimulates interest in novel
theoretical, foundational, and~mathematical studies on this inequality.
In~particular, it is important to find the quantum bound, the~analog of the
Tsirelson bound, for the original Bell inequality. It~was well-known that in
the two-qubit singlet case this bound equals $3/2,$ see, e.g.,~\cite{L8, L7}%
. A~year ago, I. Basieva~and A. Khrennikov~came with the {conjecture}~\cite%
{L17} that the same upper bound holds in case of arbitrary two-qudit states
and qudit observables coupled by perfect correlations/anticorrelations. The
question of quantum upper bound for the original Bell inequality became
actual in connection with studies on quantum-like modeling of psychological
behavior, see~related paper~\cite{ARX}.

In the present article, we~have proven this conjecture for all two-qubit
states and all traceless qubit observables and all two-qubit states and spin
qutrit observables. This~is the first step towards justifying this
conjecture for an arbitrary two-qudit case, and~the authors of the present
paper plan to continue studies on this problem. {Since in the
multi-dimensional case} the analytical expressions are very complex, it~may
be useful to try to perform preliminary numerical study, cf.~\cite{FK}.
We~also point to technique for evaluation of the quantum upper bound which
was elaborated in~\cite{FK1, FK2} and tested on the CHSH-like inequalities.
In~principle, this technique can be applied to the original Bell inequality.


\section*{Acknowledgment}

E.R. Loubenets~was supported within the framework of the Academic Fund
Program at the National Research University Higher School of Economics (HSE)
in 2018-2019 (grant N 18-01-0064) and by the Russian Academic Excellence
Project ``5-100''.

\section*{Appendix A}

\label{appendix A} {Consider the proof of Proposition} \ref{p1}.

Let, for~a joint measurement $(a_{2},b_{1})$, the~perfect anticorrelation (%
\ref{14_3}) be fulfilled and this scenario admit an LHV model (\ref{11_}).
This and (\ref{13_}) imply:%
\begin{equation*}
0\leq \int\limits_{\Omega }\left\vert f_{a_{2}}(\omega )+f_{b_{1}}(\omega
)\right\vert \text{ }\nu (\mathrm{d}\omega )
\end{equation*}%
\begin{equation*}
=\int\limits_{\Omega }\left\vert \sum_{\lambda _{a},\lambda _{b}}\left(
\lambda _{a}+\lambda _{b}\right) P_{a_{2}}(\lambda _{a}|\omega
)P_{b_{1}}(\lambda _{b}|\omega )\right\vert \nu (\mathrm{d}\omega )
\end{equation*}%
\begin{equation*}
\leq \int\limits_{\Omega }\sum_{\lambda _{a},\lambda _{b}}\left\vert \lambda
_{a}+\lambda _{b}\right\vert \ P_{a_{2}}(\lambda _{a}|\omega
)P_{b_{1}}(\lambda _{b}|\omega )\nu (\mathrm{d}\omega )\leq 2\sum_{\lambda
_{a}\neq -\lambda _{b}}P_{(a_{2},b_{1})}(\lambda _{a},\lambda _{b})=0.
\end{equation*}%
%
%
%
%

Thus, under condition (\ref{14_3}) on scenario joint probabilities, the~LHV
functions\ $f_{a_{2}}(\omega )=-f_{b_{1}}(\omega ),$ $\nu $-$a.e.$ on $%
\Omega $. Quite similarly, for~the case of perfect correlations (\ref{14_1})
we derive \ $f_{a_{2}}(\omega )=f_{b_{1}}(\omega ),$ $\nu $-$a.e.$ on $%
\Omega $. These~relations and the number inequality 
\begin{equation*}
\left\vert x-y\right\vert \leq 1-xy,\text{ \ \ }\forall \text{\ }x,y\in
\lbrack -1,1],
\end{equation*}%
give:%
\begin{equation*}
\left\vert \langle \lambda _{a_{1}}\lambda _{b_{1}}\rangle -\langle \lambda
_{a_{1}}\lambda _{b_{2}}\rangle \right\vert \pm \langle \lambda
_{a_{2}}\lambda _{b_{2}}\rangle
\end{equation*}%
\begin{equation*}
=\left\vert \int\limits_{\Omega }f_{a_{1}}(\omega )f_{b_{1}}(\omega
)-f_{a_{1}}(\omega )f_{b_{2}}(\omega )\text{ }\nu (\mathrm{d}\omega
)\right\vert \pm \int\limits_{\Omega }f_{a_{2}}(\omega )f_{b_{2}}(\omega
)\nu (\mathrm{d}\omega )
\end{equation*}%
\begin{equation*}
\leq \int\limits_{\Omega }\left\vert (f_{b_{1}}(\omega )-f_{b_{2}}(\omega
))\right\vert \nu (\mathrm{d}\omega )\pm \int\limits_{\Omega
}f_{a_{2}}(\omega )f_{b_{2}}(\omega )\text{ }\nu (\mathrm{d}\omega )\leq 1.
\end{equation*}%
This proves the statement.


\begin{thebibliography}{99}
\bibitem{L12} For the general framework on unconditional Bell inequalities
for arbitrary numbers of settings and outcomes at each site, see \cite{L13}.
General bounds on the maximal quantum violation of Bell inequalities by an $%
N $-partite quantum state have been introduced in \cite{L14, L15}.
\end{thebibliography}
\end{document}